\documentclass[12pt]{article}

\usepackage{graphicx}
\usepackage{latexsym}

\begin{document}

\title{ General Relativity Problem of Mercury's Perihelion Advance Revisited}
\author{Anatoli Andrei Vankov\\        
{\small \it IPPE, Obninsk, Russia; Bethany College, KS, USA;  anatolivankov@hotmail.com}}

\date{}

\maketitle

\begin{abstract}

The work is devoted to the  critical analysis of theoretical prediction and astronomical observation of GR effects, first of all, the Mercury's perihelion advance. In the first part, the methodological issues of observations are discussed including a practice of observations, a method of recognizing the relativistic properties of the effect and recovering it from bulk of raw data, a parametric observational model, and finally, methods of assessment of the effect value and statistical level of confidence. In the second part,  the Mercury's perihelion advance and other theoretical problems are discussed in relationship with
the GR physical foundations. Controversies in literature devoted to the GR tests are analyzed. The unified GR approach to particles and photons is discussed with the emphasis on the GR classical tests. Finally, the alternative theory of relativistic effect treatment is presented.



\end{abstract}

\maketitle

\section{Introduction}

\subsection{History in brief}

The history of the problem of Mercury perihelion advance is presented in \cite{Roseveare}, \cite{Pais} and elsewhere. Basically, it started with the work by the French astronomer Urbain Le Verrier  (1859) who calculated a perturbation effect caused by interaction of  Mercury with other planets and found {\em the anomaly}, -- the perihelion  advance unexplained by the Newton's theory. He thought that the effect could be explained by planetary matter, though it was not confirmed later on. The American astronomer Simon Newcomb \cite{Newcomb} (1882) verified the effect and assessed its value being about $39''$ (arc-seconds) per century. He also produced detailed tables of planet orbits (ephemerides) and suggested that a solution of the perihelion problem could be some alteration to the Newtonian law. Many great scientists before and after Einstein's work (1915) came to the scene with attempts to develop different approaches to the problem and improve observational techniques and calculational (perturbative) algorithms, among them the American astronomer and dedicated researcher in the problem of Mercury perihelion advance, Gerald Clemence.  

In a long run (from Le Verrier to Einstein), there was a wide range of admissible  values of ``the anomaly'' (say, from 5 to 50 arc-seconds per century). Probably, a strive for ``a true explanation'' of the anomaly prevailed over work on determination of ``the true anomaly'' value from astronomical observations, still of poor precision.  With Einstein's GR advent, however, hot discussions of the problem were not ceased. Many Einstein's prominent counterparts  remained highly critical about  the Einstein's theory or did not accept it at all. When Einstein received a Nobel Prize (1922) for great achievements in Physics, there was a special note saying that the prize was given ``without taking into account ... relativity and gravitational theories'' \cite{Pais}.  

One should bear in mind that in the historic time till, say, mid 50s, a conduction of computations of planetary orbits with mechanical or electromechanical calculators was an extremely hard job. The electronic computer era began in about 1945, while the state-of-art maturity came after Clemence's activity discontinuation. In spite of technical difficulties, Clemence made a tremendous contribution to the problem study (early 40s till late 60s), especially in matching  physical theories and astronomical observations and achieving better clarity of the problem and higher precision of data analysis.  He rechecked Le Verrier's and Newcomb's results and updated planetary data  \cite{Clem1} (1943), \cite{Clem2} (1947), \cite{Clem3} (1948), \cite{Clem4} (1965) . It was clearly understood by Clemence (and pretty much so by his predecessors) that the accuracy of the assessment of ``the anomaly''  is restricted by its smallness (less than 1/100 of ``perturbation background''). There were unavoidable methodological problems such as a connection of time scales in different observations as well as determination of inertial reference frames the effect must be referred to. Basically, he stated that, given realistic uncertainties of input data and model parameters, the planetary ephemirides can be, in principle, adjusted by minimization of the anomaly gap in concordance with the General Relativity prediction. This was a solid result but still not a resolution of the problem because of a number of controversial issues remained unresolved. In 1947, Clemence noted  \cite{Clem2} (1947) : 

``According to general theory of relativity, the elliptical orbit of a planet referred to a Newtonian frame of reference rotates in its own plane in the same direction as the planet moves... The observations cannot be made in a Newtonian frame of reference. They are affected by the precession of the equinoxes, and the determination of the precessional motion is one of the most difficult problems of observational astronomy. It is not surprising that a difference of opinions could exist regarding the closeness of agreement of observed and theoretical motions... I am not aware that relativity is at present regarded by physicists as a theory that may be believed or not, at will. Nevertheless, it may be of some interest to present the most recent evidence on the degree of agreement between the observed and theoretical motions of the planets''. 

The situation in observational astronomy radically changed with the technological boost started in 1950s and continued since then. After the WW2, astronomers armed with advanced observational tools and computer technologies took a chance to pursue ambitious unprecedented projects in observational astronomy and celestial mechanics, such as the international celestial reference frame and coordinate system, the Earth rotation service with coordinate system transformations, the database for a wide range of astronomical and physical constants, parameters of theoretical N-body gravitational models, etc. 

The initiative was launched to include the GR theory in the form of the Parameterized Post Newtonian (PPN) approximation. It was motivated by needs Astronomy and Astrophysics, space missions and other national tasks related to time and position precision in different reference frames. Such tasks could be fulfilled only with the use of contemporary electronic technologies. In particular, it became possible to study errors in approximate solutions of N-body problem and the role of relativistic corrections there. This is what we have today under IAU scientific governance over ephemerides centers such as the Jet Propulsion Laboratory (JPL), USA, Institute of Applied Astronomy, Russia, and some others. 

As for  the ``anomaly'' problem, we are not saying that much better clarity than at Clemence's time has been achieved. Some astronomers have continued the anomaly studies basically with the old methodology, which includes, on the one hand, a calculation of the Mercury's perihelion advance with ``perturbing'' forces from all other planets; on the other hand, a comparison of theoretical results with empirical data related to Mercury's motion in order to fit theoretical ephemerides as close as possible to the Einstein's predicted value.

\subsection{PPN and current status of the perihelion advance problem}

Work on GR (PPN) extension to the N-body problem was actually started by Einstein with coauthors \cite{EIH} (1949) and was continued during Clemence's time and later on, see, for example,  \cite{Weinberg} (1972), \cite{Misner} (1973), \cite{Soffel} (1989), \cite{Brumberg} (1991)), \cite{Clifford} (1993), \cite{Moyer} (2000),   \cite{SoffelIAU} (2003), \cite{Pitjeva1} (2005), \cite{Pitjeva2} (2005), \cite{Kopeikin} (2006),  \cite{Klioner} (2007),   \cite{Muller} (2008), and references there. 

In the  70s and later on, the N-body problem was theoretically formulated in the parameterized 
post Newtonian (PPN) formalisms at the level of ephemerides calculations. The GR perihelion advance problem is supposed to be included in the formalism. As well known, the GR equations does not provide exact solutions for most of practical problems, the N-body system, in particular. The PPN approximation idea is to linearize the equations under weak-field conditions for approximate N-body solutions. Inevitably, the relativity essence such as a concept of proper-versus-improper quantities has to be sacrificed. 

Our criticism of the PPN methodology is expressed in the question: approximation to what? Definitely, it is not the approximation of the exact GR solution of N-body problem, because such a solution does not exist. This is why the existence of the PPN frame with the corresponding coordinate system cannot be justified, it is postulated. In the special important case, $N=1$ plus a test particle, the exact solution does exist; this is the Schwarzshild metric, which is valid in the entire range of field strength. In PPN methodology, the Schwarzshild metric is replaced with the approximate (weak-field) solution containing the PPN parameters $\beta$ and $\gamma$. The PPN formalism is intended to account for perturbation of the planetary system due to planet interactions, basically, with the use of Le Verrier's method of Newtonian mechanics. Nowadays, however, the N-body problem has an exact computer-supported solution in the Newtonian model.  

The incorporation of the PPN formalism into the ephemerides systems took place in a process of continual ``fitting'' of ephemerides systems to observation database. Consequently, the PPN parameters were (allegedly) well pinned. Overall, an optimistic picture is claimed that astronomical constants and planetary ephemerides are adjusted to hundreds of thousands of different observations (astronomical, radiometrical and others). Allegedly, any orbit can be accurately (in a sense of ``fitting'') displayed in a desirable coordinate system and in time directed into past or future for a long enough period. We put the word ``allegedly'', because there is a difference between ``fitting'' and ``evaluation of statistical significance of observational data'', -- the issue discussed later.

\begin{table}
\center\caption{\label{Table} {\bf Sources of Mercury perihelion precession}  }
\begin{tabular}{ll} 
\hline
{\bf Amount (arcsec/ century)} & {\bf Cause} \\
\hline
5030 &  Precession of the equinoxes \\
\hline
530  &  Perturbation by planets \\
\hline
43   &  General Relativity \\
\hline
5603 &  Total   \\
\hline
5600 &  Observed   \\
\hline
\end{tabular}
\end{table}

   A practical coincidence of prediction and observation is demonstrated in Table \ref{Table} (the data  in literature usually referred to). Here, we omit originally posted precision numbers. In terms of relative accuracies, they are $10^{-5}$ for the equinoxes precession (noted by Clemence as the biggest problem), and  $10^{-3}$ for the predicted GR effect.  

In \cite{Pitjeva2} (2005) we read, for example: 
 ``As the uncertainties in
 [ephemerides] parameters decrease, the domain of possible
values of the relativistic parameters narrows, imposing
increasingly stringent constraints on the theories
of gravitation alternative to General Relativity.'' 

The above and many similar statements should be understood as a claim of a huge leap achieved after Clemence's time in Astronomy with tremendous consequences in Physics. There is a firm consensus among astronomy and physics communities, mass media as well, that the perihelion advance test is the accomplished task. We think, however, that physical reality is more complicated, and the above  precision numbers for physicists must seem to be to a great extent fictitious.

As a result of our investigation of the problem, we come to the conclusion that the long-standing efforts (starting from Le Verrier till present days) to confirm the Einstein's prediction of the perihelion advance have not brought physical evidence of the confirmation in rigorous terms of physical and statistical theories. Further, we present arguments concerning two sides of the problem: 

1)  Astronomical observations and their ``matching'' with the Einstein's prediction; 

2)  Theoretical rigor of the Einstein's prediction.


\section{Calculation versus Observation}

\subsection{What and how it can be observed}

Let us see the example of Mercury's ephemerides calculations in the geliocentered frame, Narlikar and Rana \cite{Nar1} (1985), Rana \cite{Nar2} (1987), Fig.\ref{RaNa}. A classical motion of Mercury's perihelion advance caused by the interaction of planets gives the effect 530 arc-sec/per century, see Table \ref{Table}.  This is a fluctuating part of the background, on which the GR predicted effect of rate of 43$''$ per century must emerge.

In the authors' works commented below, a powerful (Burlirsch-Stroer) numerical method of differential equation solution is used with the time-step about one day chosen to provide a calculational (accumulated in integration) error of one part in $10^{13}$. Such a precision is required to ensure a stability of numerical integration over about 300 year period and reverse back. Changing the step from 1 day to two days results in a computational error about 0.001 arc sec. At the same time, initial conditions are known not better than seven significant digits. (A position and a velocity of a center of every planet at a given instant are meant). This must translate into a cumulative error in calculations about 10 arcsec per century. In order to maintain numerical stability of calculations, the authors have to control more significant digits. 

As we understand, the computational ``instability'' in lower order figures reflects physical reality of the system of interacting planets (as in Fig.\ref{RaNa}). A small perturbation acquires an ability to propagate into nearby area and inflate in time. Indeed, a motion of every planet has many degrees of freedom so that an orbit can differ from the elliptic one in however many modes. The authors do not conduct calculations in the relativistic (PPN) model because of a drastic rise of computer time. This is quite unfortunate because one would like to see in the PPN formalism the effect of ephemerides fluctuation and time instability in the determination of a small regular relativistic effect on top.

\begin{figure}[t] 

\includegraphics[width=5in, height=2.8in]{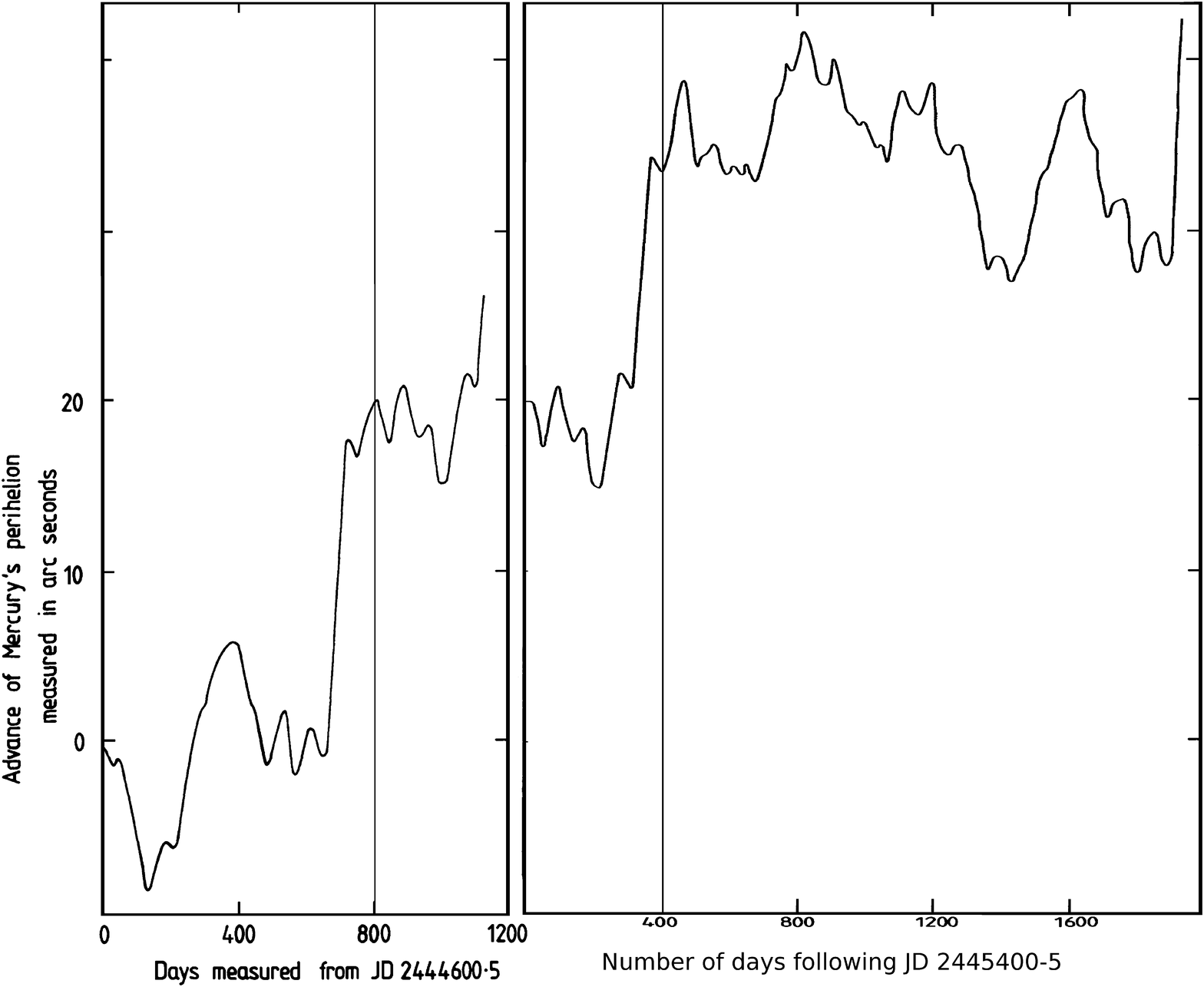}
\label{RaNa}
\caption{\label{RaNa}   {\small Position of heliocentric longitude of perihelion of Mercury with time. Vertical scale: cumulative effect in arc-seconds. Horizontal scale: days past starting from December 27, 1980 till April 10, 1984 (left picture \cite{Nar1}),  and March 7, 1983 till February 9, 1988 (right picture \cite{Nar2} (1987)). It is seen that the graph for the last 400 days in the left picture and the graph for the first 400 days in the right picture overlap in time (as indicated by vertical lines). Total number of Mercury's orbital periods is 25. The authors' comments are given in the text.} }

\end{figure}


The question arises how to determine the difference between  ``calculation'' $C(T)$  and ``observation'' $O(T)$ for different instants $T$ in view of fluctuation of angular motion. In practice, observational data are treated  with the use of polynomial ``averaging'' interpolation such as 
$\Gamma_1 (T)$ with the term $5600$ $\cdot$ $T$ and $\Gamma_0 (T)$ with the term $530$ $\cdot$ $T$, so that one has to deal with the problem of reduction  $\Gamma_1 \to \Gamma_0 $. The authors note that an uncertainty about $\pm 1''$ of geocentric observations translates into the GR effect precision about 10 $\%$.  From the authors' work it follows that the fine structure in calculations and observations must be accurately accounted for in order to preserve true physical meaning of measured quantities.

The important part of their works is a criticism of the method of  ``Einstein's effect fitting'' used by  Clemence and other authors in later times. In particular, the results obtained by Clemence, also by Morrison and Ward \cite{Morrison} (1975), are discussed. The matter is that the Clemence's results were reanalyzed  in  \cite{Morrison} in view of the fact that precision of data for masses of planets and time scalings were significantly improved since Clemence's time. This noticeably changed the $C(T)$ while the $O(T)$ was only slightly corrected. Nevertheless, the fitting (least square) criterion remained excellent.  The conclusion of Rana's work \cite{Nar2} (1987) is given below.

``The low precision of the geocentric angular data having an error of 1$''$ are incapable of giving the rate of motion of the perihelion of Mercury to better that 3$''$ per century. Hence the determinations apparently good to 0.3$''$ per century are spurious. Using too low a rejection level only those data that happen to agree with the initially assumed value of the rate of motion of the perihelion contribute, and one will always obtain a very small correction to the initially assumed value. So all the existing observational estimates, namely those made by Clemence, Morrison and Ward, and the JPL groups, are suspect.''

The author noted that the ephemerides uncertainties must be increased substantially. Similar opinion was expressed by Pitjeva \cite{Pitjeva1} (2005). The different approach to the problem was made in \cite{Takeshi} (1993), when the least square criterion was evaluated without paying attention to the Einstein's predicted value. After discarding few points, which looked suspicious, the author found a drastic drop in the least square number for the advanced perihelion value about 15-16$''$ (instead of 43) per century. His comment was:

``The only conclusion one can draw from the data is thus that they do not contribute to a
decision as to whether the actual motion of the ascending node of the orbit of Mercury
exceeds that predicted by the theory''.

\subsubsection{What has been observed}

Now we state that the Mercury's relativistic effect has never been directly observed and even not evaluated from circumstantial astronomical evidence. The matter is that the GR theory, at least as it given in literature, does not provide a clue about distinguishing between the classical drag along with the equinoxes  precession, on the one hand, and relativistic effect, on the other hand. There is no other way but look for an admissible anomaly gap to be filled with the predetermined perihelion advance of 43$''$ per century as tight as possible, no matter of what kind the effect is. Such ``a gap fitting'' cannot be termed ``the confirmation of the GR prediction''.

Another issue is a statistical meaning of the gap fitting. There is no single publication devoted to the treatment of observations of Mercury perihelion advance; the claimed numbers are stated in different works on empirical data not treated in rigorous terms of statistical theory. A bad fitting practice and the precision concept abuse should be noticed. At the same time, the usage of standard ``precise'' initial conditions in ephemerides calculations makes the results stable what creates an illusion of their high-precision, while their real precision remains unknown.  

To avoid any terminological ambiguity and fruitless disputes on this important issue, we present a brief review of the statistical method that is usually used in physical experimental studies and, in our opinion, should be used in the perihelion advance investigation as well (see the $Appendix A$ with comments). In the following sections, we investigate theoretical rigor of the GR predictions.


\section{Classical and GR orbits}


\subsection{Classical basic equations}

Before making a comparison of classical (Newtonian) and relativistic (GR/SR) formulations of the problem, let us discuss the classical equation of motion for a point particle in a spherical symmetric field \cite{Goldstein}. The conserved total energy $\epsilon_0$ in the dimensionless form is given by  
\begin{equation}
\epsilon_0=1-\frac{r_g}{r}+ \frac{1}{2}\beta_r^2+ \frac{1}{2}\frac{l_0^2}{r^2}
\label{1}
\end{equation}
where the conserved angular momentum is $l_0 =r\beta_\theta=r^2(d\theta /dt)$; $r_g=G M$, $\beta_r =(dr/dt)$, the speed of light at infinity $c_0=1$. One can replace the equation (\ref{1}) with the almost equivalent one having an ``relativistic appearance'' of the squared total energy under the weak field conditions: 
\begin{equation}
\epsilon_0^2=1-2\frac{r_g}{r}+ \beta_r^2+ \frac{l_0^2}{r^2} 
\label{2}
\end{equation}
Here, the second-order terms are neglected, one of them being $r_g l_0^2/r^3$, similar to that responsible for perihelion advance in GR.

With the use of relationship $\beta_r= (dr/du)(du/d\theta)\omega_\theta$,   $\omega_\theta=l_0/r^2$ with $u=1/r$, the equation in the standard form is given by
\begin{equation}
(du/d\theta)^2=(\epsilon_0^2 -1)/l_0^2 + 2(r_g/l_0^2) u - u^2
\label{3}
\end{equation}
In the traditional form  in polar coordinates, the equation has two parameters: the eccentricity $e \ge 0$, and the so-called semi-latus rectum $p$ (the distance from a focus to the ellipse):
$e^2=1+(\epsilon_0^2-1)r_g^2 /l_0^2$; \ $p=l_0^2/r_g$,
\begin{equation}
(du/d\theta)^2=(e^2 -1)/p^2 + 2 u/p - u^2
\label{3b}
\end{equation}
with the roots:
\begin{equation}
u_1=(1-e)/p, \  \  \  u_2=(1+e)/p 
\label{3a1}
\end{equation}
leading to
\begin{equation}
\theta_2-\theta_2=\int_{u_1}^{u_2} {du \left[(u-u_1)(u_2-u) \right]^{-1/2}}
\label{3a2}
\end{equation}
\begin{equation}
r(\theta)= p_0/\left[(1+e\ \sin(\theta-\theta')\right]
\label{4}
\end{equation}
where $\theta'$ can be fixed in the initial conditions.

\subsection{Orbit classification: $e$-criterion versus $\sigma$-criterion}

The traditional classification of orbits is based on the $e$-criterion:

$e>1$,\ \ \ $\epsilon_0>1$,  \ \ \ \ \ \ \ \ \ \ \ \ \ \ hyperbola;

$e=1$,\ \ \ $\epsilon_0=1$,   \ \ \ \ \ \ \ \ \ \ \ \ \ \ parabola; 

$e<1$,\ \ \ $\epsilon_0<1$,  \ \ \ \ \ \ \ \ \ \ \ \ \ \  ellipse;

$e=0$,\ \ \  $\epsilon_0=1-r_g/2r_0$,\ \ \  circle.

\medskip

The above classification, though customary in many applications, makes a study of the relativistic perihelion problem practically impossible. The two parameters $p$ and $e$ are  combinations of physical quantities, as they historically came from geometrical studies of curves. 
In Physics and Astronomy, one has to deal with a family of orbits specified by one or two physical parameters. In this work, we specify the initial perihelion condition: $\beta(r_0)=\beta_0$, at $r=r_0$, $(\theta-\theta')=0$ with $l_0=r\beta_\theta=r_0\beta_0$. 

Further we are going to exploit a non-conventional family of one  parameter $\sigma = r_g/r_0\beta_0^2$ for the equation of motion with a radial function being $\xi(\theta)=r_0 u(\theta)$. In studies of relativistic perihelion advance, a consideration of motion in terms of the azimuth angle $\theta$ was found quite useful. Thus we introduce the classical equation in the nonconventional form:
\begin{equation}
(d\xi/d\theta)^2=(1-2\sigma) + 2\sigma \xi - \xi^2
\label{5}
\end{equation}
It is a classical harmonic equation, which can be presented in different forms, for example
\begin{equation}
d^2 \xi /d\theta^2=\sigma -\xi
\label{6}
\end{equation}
or, by a substitution $\xi - \sigma =x$:
\begin{equation}
d^2 x/d\theta^2 = -x
\label{7}
\end{equation}
The solution to (\ref{5}) may be chosen in the form
\begin{equation}
\theta=\cos^{-1}{\frac{(\xi-\sigma)}{(1-\sigma) }}
\label{8}
\end{equation}
\begin{equation}
\xi(\theta)= r_0/r(\theta)=\sigma + (1- \sigma) \cos(\theta) 
\label{9a}
\end{equation}
or
\begin{equation}
r(\theta)/r_0= \left[\sigma + (1- \sigma) \cos(\theta)\right]^{-1} 
\label{9b}
\end{equation}
As wanted, the equation is left with only one parameter $\sigma$, which absorbs $r_g$, $r_0$, and $\beta_0$. The parameter is intimately related to the 3d Kepler's law.


Thus, we suggest the following $\sigma$-classification of orbits illustrated also in Fig.\ref{FiveOrbits}.

$0< \sigma < 0.5$, \ \ \ \ \ \ hyperbola;

$\sigma=0.5$,  \ \ \ \ \ \ \ \ \ \ \ \ \  parabola; 

$0.5 < \sigma < 1$,  \ \ \ \ \ \  overcircle ellipse;

$ \sigma = 1$,\ \ \ \ \ \ \ \ \ \ \ \ \ \ \ circle;

$1 < \sigma < \infty $. \ \ \ \ \ \ \ subcircle ellipse.

In accordance with the $\sigma$ criterion, eccentric orbits can be of two types. The ``overcircle'' orbits are characterized by initial kinetic energy $\epsilon_k$ and the total energy $\epsilon_0$ being greater that in the circular motion with $r=r_0$, while the speed at apohelion is $\beta_a <\beta_0$. For the ``subcircle'' orbits it is just opposite: $\epsilon_k$ and $\epsilon_0$ are less than that in the circular motion, while $\beta_a >\beta_0$. The subcircle motion includes orbits with $\sigma \to\infty$, as  $\beta_0\to 0$, when the particle's speed near the point source becomes however high. This case must be considered in the relativistic rather than classical framework.

\begin{figure}
\includegraphics[scale=0.75]{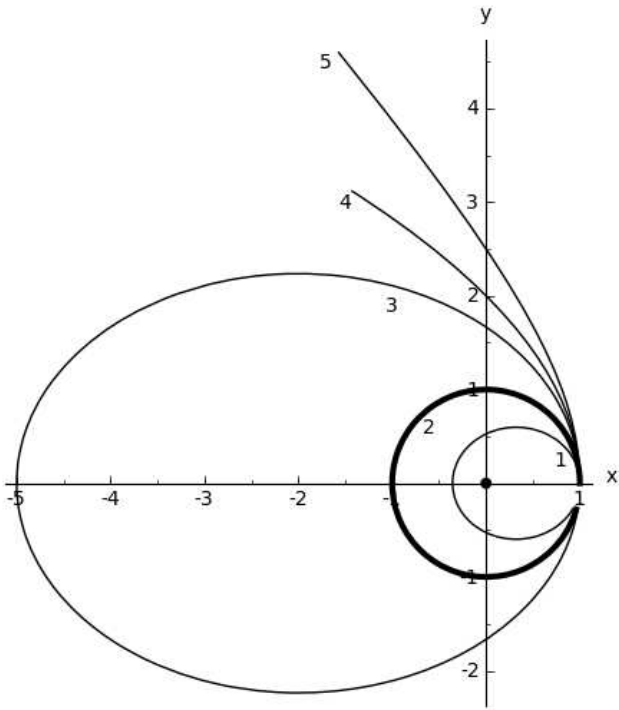}
\label{FiveOrbits}
\caption{\label{FiveOrbits}  Example of $\sigma$ family of classical orbits. 1. Subcircle ellipse, $\sigma$=1.9; 2. Circle (thick line), $\sigma$=1; 3. Overcircle ellipse, $\sigma$=0.6; 4. Parabola, $\sigma$=0.5; Hyperbola, $\sigma$=0.4. The gravity center is at the coordinate origin. All orbits are produced by launching a point test particle at the perihelion point $x$=1 with the initial speed   $\beta_0=\sqrt{{r_g}/\sigma}$. }
\end{figure} 

In a family of orbits of different radii $r_0$ (like in a Solar planetary system) it is convenient to use the $\sigma$-classification for the pair of variable parameters,  $r_0$ and $\beta_0$, with $r_0$ being the perihelion (minimal) distance from a focal point. In this case, the notion of ``subcircular'' orbit is not relevant.

The $e$-criterion cannot, in principle, distinguish between the two types of ellipses (overcircle and subcircle) because it suggests variations of physical parameters all at once with $e$. If  $e$ varies with $p$ fixed, the family consists of orbits with gravity centers sliding along the major axis, what makes a physical analysis difficult. Consider, for example, an orbit with a small eccentricity. The equation (\ref{3b}) has two roots: 
$u=u_0(1\pm e)$, or $r=r_0(1\mp e)$.  Physically this makes a false picture of a point particle being at the same time in orbital motions and in a state of radial harmonic oscillation in the parabolic well. 

In reality, the particle undergoes an asymmetric periodic motion with respect to the gravity center between a perihelion point $r=r_0$ of a lowest potential and the apohelion point of a maximal potential. The correct physical solution comes from (\ref{5}). The first root is $\xi_1=1$, or $r_1=r_0$. The second root is $\xi_2=2\sigma -1$. For $\sigma=1 \mp \Delta\sigma$, and small $\Delta\sigma$, it is $r_2=r_0(1\pm\Delta\sigma)$. This gives a subcircle or overcircle ellipses. 

The example of two roots in the $\sigma$ versus $e$ parametrization of orbits is important in view of the idea known from literature about the assessment of the GR perihelion advance. It suggests that the orbit slightly different from the circular should be analyzed on subject of its angular and radial frequencies with respect to the circle frequency. The radial oscillation is expected to be the harmonic one in the parabolic potential well. It is thought that by finding the difference between angular and radial frequencies one could find the resulting perihelion advancement. As shown above, this is a wrong approach. We shall return to this issue later in more details. In our further analysis of the conditions providing the relativistic perihelion advance, the $\sigma$-criterion in perturbations of a circular orbit is used.

\subsection{GR Schwarzschild Dynamics of particle and photon}

GR Physics of a test particle and a test photon is described by the Einstein's field equations  \cite{Einst1}, \cite{Bergmann}, or equivalently, the Schwarzschild metric \cite{Schwarz}, \cite{Taylor}. The latter in polar coordinates is given by:
\begin{equation}
ds^2= dt^2/\Gamma_r^2 - \Gamma_r^2 dr^2 - r^2d\theta^2
\label{Schwa}
\end{equation}
where we denote $\Gamma_r^2=1/(1-2 r_g/r)$, $c_0=1$, and $ds=d\tau$ ($c_0=1$). The  equation is obtained from the Hamilton's variational principle for the Schwarzschild field, \cite{Pauli}, \cite{Fock}.
\begin{equation}
\delta \int {ds}=\delta \int{L dt}= 0, \  \  \ L=d\tau /dt
\label{VarPri}
\end{equation}
with the following integrals of motion
\begin{equation}
\epsilon=(dt/d\tau)/\Gamma_r^2
\label{conser1}
\end{equation}
\begin{equation}
l_0=r^2(d\theta/d\tau)=r\beta_\theta
\label{conser2}
\end{equation}
Here, $\epsilon$ is treated as the total energy, and $l_0$ -- the angular momentum. They are readily follow from the time symmetry and the space isotropy in a manner of classical mechanics.
The third equation is a definition of the proper time (\ref{Schwa}) usually given by
\begin{equation}
(dt/d\tau)^2/\Gamma_r^2-(dr/d\tau)^2 \Gamma_r^2- r^2 (d\theta/d\tau)^2= \kappa
\label{tau}
\end{equation}
Here $\kappa=1$ for the particle, and  $\kappa=0$ for the phton.

The first two expressions are valid for the particle and the photon. The equations of motion of the particle and the photon in the Schwarzschild field follow  from (\ref{conser1}), (\ref{conser2}), (\ref{tau}). The term which causes the GR effect (further called the GR-term) is inclosed into the framed box. The equations  will be used in our further analysis of GR effects. Below, the orbit equations are given

1. Particle (planet)

\begin{equation}
(du/d\theta)^2=-(1-\epsilon_0^2)/l_0^2 + 2(r_g/l_0^2) u - u^2  +   
\fbox{  \parbox{1.1cm}{$  2 r_g u^3 $ }  }
\label{Ein1}
\end{equation}

\medskip

2. Photon (light ray)

\begin{equation}
(du/d\theta)^2=\epsilon_0^2)/l_0^2  - u^2  +   
\fbox{  \parbox{1.1cm}{$  2 r_g u^3 $ }  }
\label{pho}
\end{equation}

\subsection{Introduction of the GR-term in the particle orbit equation}

In classical mechanics the ``effective'' potential energy is
\begin{equation}
V(r)=\epsilon^2-\beta_r^2 = 1-2r_g/r+ l_0^2/r^2
\label{clasV}
\end{equation}
where $\beta_r^2$  is the radial part of kinetic energy $\beta^2=\beta_r^2+\beta_{\theta}^2$.
The extremum equation gives a circular orbit with the radius $r_c$
\begin{equation}
r_c=l_0^2/r_g=r_g/\beta_0^2, \  \  \    \  \sigma=r_c/r_c\beta_0^2=1
\label{clascir}
\end{equation}

The GR perihelion advance is associated with the GR-term $(r_g l_0^2/r^3)$, which is of the second order of smallness $(r_g l_0^2/r^3)\sim{(r_g/r)^2}$ in the energy balance. At this point, we do not comment the fact that the GR problem is formulated in terms of proper time $\tau$. It should be also noted that the GR theory is supposed to explain laws of gravitational physics exclusively in terms of curved spacetime; the concepts of potential and kinetic energy are  not in the GR arsenal. 

It is habitual to consider the GR-term  in the ``GR effective potential energy'' to be compare with the classical case (\ref{clasV}: 
\begin{equation}
V_{ef}(r)=\epsilon^2-\beta_r^2 = 1-2r_g/r+ l_0^2/r^2 -
\fbox{  \parbox{1.4cm}{$  2r_g l_0^2/r^3 $ }  }
\label{term1}
\end{equation}
In the angular part $\beta_{\theta}^2$, the quantity $l_0=r_0\beta_0$ must be the conserved orbital momentum.  

The equation (\ref{term1}) can be rewritten
\begin{equation} 
 V_{ef}/\beta_0^2 =1/\beta_0^2 -2\sigma\xi +\xi^2 - \fbox{  \parbox{1.6cm}{$  2 (r_g / r_0) \xi^3 $ } } 
\label{term2}
\end{equation}
Finally, it can take the form of equation of orbital motion, by analogy with (\ref{5})
\begin{equation}
(d\xi/d\theta)^2=(1-2\sigma) + 2\sigma \xi - \xi^2 + \fbox{  \parbox{1.7cm}{$ 2(r_g/r_0) \xi^3 $ } }    
\label{term3}
\end{equation}
The above GR equations have common extrema and can be used for determination of orbital motion parameters. The extrema are readily found by equating the derivative of right side of equation to zero. 

Let us see potential energy curves with and without the GR-term for a strong field condition in order to understand the role of the GR-term.
The graphs (\ref{term1}) for the classical potential energy  (upper line, no GR-term) and its GR version (below, with the GR-term) are shown in Fig.\ref{3roots}. Both curves are plotted for $l_0^2=20$ in $r_g^2$ units. Radii are given in units of $r_g$.  Consequently, the classical radius of circular motion is $r_c=20$ and the corresponding speed $\beta_0=0.22$. For the GR curve the radius of the circular motion is about  $r_0=16.8$.
The ``peak'' at the radius $r_1=3$ is caused by the GR-term. 

Here is ``an inner'' radius about $r^*=2.5$, which, in our opinion, is a fictitious solution. One can consider it a solution for a confined particle in a separate circular orbit slightly above the so-called Schwarzschild sphere with a speed of motion close to $c_0$. If to take $r_g=1.5\cdot 10^3$ m (the gravitational radius of the Sun), the orbital radius will be $R=2.3\cdot 10^4$ $m$. Compare it with Mercury parameters: semi-latus rectum $p=5.8\cdot10^{10}$ $m$, $\beta=1.3\cdot10^{-4}$. Looking at the picture, one has to bear in mind the GR assertion that the GR-term drives the perihelion precession in the direction of motion.

\subsection{Classical versus GR treatment of the GR-term}

The GR-term cannot arise in the Newtonian orbit equation except for usage of sticks and ropes.  Having this said,  one can treat the equations {\em with the GR-term} classically (consistently with the criterion $\sigma$),  as in (\ref{term3}). For  $r_0=r_c(1-3r_g/r_0)$,  we have $\sigma=r_g/r_c (1-3r_g/r_c)\beta^2 >1$ that shows a subcircle ellipse. It can be transformed into a circle by variation of $\beta$: with $\beta_1=\beta_0 (1+3r_g/r_0)$, we have  $\sigma=1$. 

In  our approach, the cubic algebraic equation has always a unit root $\xi_{-}=1$. One can find the circular solution or the elliptic ones along with the third root $\xi^*=r_0/r^*$ to the precision of any higher degree and reduce the equation to the quadratic form. The main solution is just an ordinary classical one. One can vary physical parameters either keeping $l_0^2$ constant or redefining it. The main thing to remember is that the classical circular orbit always abide by the inertial rotation law  $\sigma=1$. 

Suppose, from physical considerations, it is known that the {\em circular orbit} is characterized by parameters, which make the criterion  $\sigma\ne 1$. This would be a sign to treat the equation of motion alternatively, say, in terms of SR or GR. Such an example is known in SR Kinematics, -- the Thomas precession, discussed later.

\begin{figure}[t]
\includegraphics[scale=0.65]{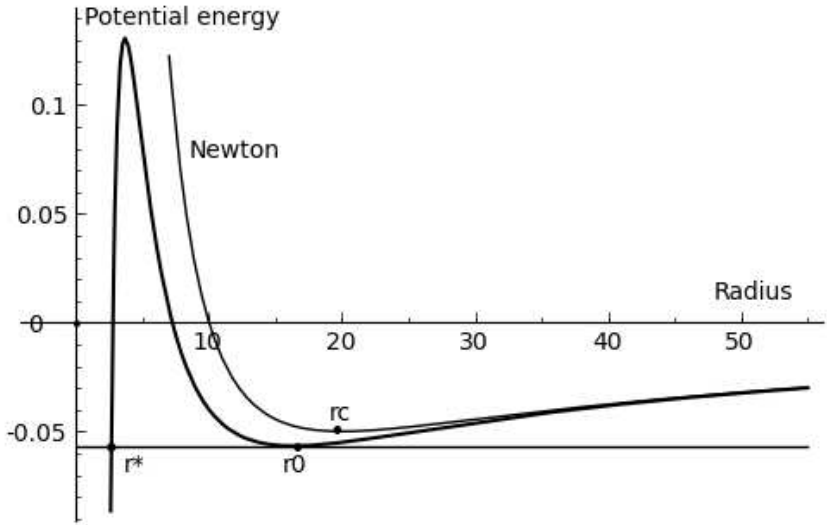}
\label{3roots}
\caption{\label{3roots} GR ``effective'' potential energy (thick line) as a function of radius in $r_g$ units. A circular motion radius is $r_0$. The upper curve is the Newtonian potential (the GR term removed)  with the corresponding radius of circular motion $r_c$. In both cases, $l_0^2=20$ $r_g^2$. Notice the shift of two graphs due to the GR-term. For comments on $r*$ and other details, see the text.}
\end{figure}

Now let us consider the approach to the GR perihelion advance problem based on the GR concept of effective potential. The latter can be ``constructed'' from the Schwarzschild metric, as discussed in the famous monograph ``Gravitation'' by Misner with co-authors \cite{Misner} (1973). They suggested that the Einstein's perihelion advance of $3 r_g/r_0$ (rad per 2$\pi$) could be considered in an approach different from the Einstein's one. The starting point is to consider an orbit with a slightly different energy than in a pure circular motion. The authors gave some hints to the solution but suggested that students would do the work. Such a work was actually done by R. Wald (1984) \cite{Wald}, and it was widely popularized later, for example, \cite{Taylor}, (2000). The assumption (mentioned above in connection with the $e$-criterion of orbits) is made that the value of perihelion advance  could be calculated as a difference between the angular frequency  $\omega_\theta$ and the frequency of radial oscillation $\omega_r$. Due to the GR-term impact, the angular frequency increases by the factor of
(1+ 1.5 $r_g/r_0$). The radial frequency must be found from the second derivative of the potential as an approximation of the law of harmonic oscillation in a parabolic potential well.  One can visualize the picture of harmonic oscillation by examining the solution $u_0\pm e$ from the $e$ classification of orbits. The radial frequency must decrease by the factor of (1+ 1.5 $r_g/r_0$). A total relative change of a classical angular frequency by a factor (1+ 3 $r_g/r_0$)  appears to be in agreement with the Einstein's prediction.

 There are objections. First, a comparison of classical and relativistic cases needs 
 relativistic-versus-classical arguments and criteria. Second, in order to make the perihelion advance in the direction of motion, the angular frequency must {\em decrease} (not increase). We shall discuss this issue later in more details. Third,  the radial and angular oscillations cannot be manipulated in arbitrary perturbations; they are engaged, as seen from relations $\beta_r \sim \omega_r \sim {dr/d\theta}\sim \omega_\theta$. Fourth, the picture of parabolic  well appears in the $e$ orbit classification, which is, though geometrically correct, physically misleading and wrong in this case, as discussed earlier.  

We think that work with the {\em relativistic} equations (\ref{term1}), (\ref{term2}), (\ref{term3}) in Newtonian environment is, in principle, a wrong idea. Einstein's original approach was technically and methodologically different. He worked with the equations embedded into the GR framework with conceptual attributes such as symmetry and conservation laws, particle and photon parallels, finally, proper versus improper time scales. Whether he took advantage of it is another matter. Anyway, one can raise the question why to search for alternatives when the solution of the perihelion advance problem was given by Einstein in 1915 and never seriously doubted. Or was it?

\section{The GR angular advance, and the SR Thomas precession}

\subsection{The SR Thomas precession}

The mentioned earlier Clemence's notice that the GR perihelion advance is a rotation of orbital plane deserves a full attention. Most likely, Clemence was aware of the SR Kinematics effect, -- the Thomas precession \cite{Thomas} (1926), which is only one sixth the GR effect but resembles it in all other respects. Can the Thomas precession contribute to the GR effect? To clarify this question and find out the ways of distinguishing between classical and relativistic motion, we need to discuss the phenomenon. 

In a rotating frame, centripetal acceleration is caused by a stress, which affects clock rates. This suggests that the Thomas precession has to be considered in terms of SR Dynamics. Nevertheless, it is often treated in terms of SR Kinematic as an approximation tolerable to the extent when dynamical corrections to the effect are neglected. Still one has to bear in mind that in SR Kinematics of uniform linear motion there are no preferred coordinate system, while a system uniformly rotating by inertia is a non-inertial, hence, preferred one.

Consider a uniformly rotating rigid frame with a coordinate system $S'$ attached to the frame. The system rotates about the origin $O'$ with the angular speed $\omega_0$. The rotation is seen from an inertial coordinate system $S$ with the center $O$ coinciding with the center $O'$ both being at at rest with respect to the stars far away. Notice,  radial scales in $S$ and $S'$ are similar. A point $r'=r_0$ is in a circular motion in $S$; it moves along the circular arc $s$ with the speed  $\beta_0=r_0\omega$. The arc path interval is $\Delta s(t)=\beta_0 \Delta t=r_0\omega \Delta t$, where $t$ is a coordinate time related to the proper time of rotating observer $t'=t/\gamma$. The Lorentz factor is $\gamma_0=(1-\beta_0^2)^{-1/2}$, hence, $s/s'=\gamma_0$. This is the SR Thomas precession phenomenon.

\subsection{The panetary version of the Thomas precession}

Let a hoop rotate {\em about the gravity center} at $O$ coinciding with $O'$ so that the $\sigma$-criterion  $\sigma'= r_g/r_0 \beta_0^2=1$ in $S'$ and, similarly $\sigma= r_g/r_0\gamma_0^2 \beta_0^2=1$  in $S$ is satisfied. In both systems, the hoop is subjected to no internal stress. For $\sigma> 1$ the stress would arise due to hoop stretching, for $\sigma< 1$ for compressing. Now, cut the hoop into small pieces, and we have the version of Thomas precession of a planet in a circular motion. The orbit will acquire an elliptic, nearly circular, form if a small instant perturbation $\delta r$ is given. Hence, the arc advance can be observed in $S$ in the form of precession of the line connecting perihelion and apohelion points. In both systems, the following definition of the half-circle period applies: the time passed along the arc $\pi r_0$ from the event ``perihelion'' to the event ``apohelion'' in a limit $\delta r\to 0$. 

A planet or satellite precession can be observed in a pure circular motion. If the clock in $S'$ ``flashes'' every period, the flash point is seen precessing in $S$ in the direction of motion. The comoving observer in $S'$ cannot see the precession unless she periodically snaps stars far away. Mathematically, the orbit precession is described by a periodic function with the argument $\omega t/\gamma\approx \nu\omega t $, where $\nu=(1-\beta_0^2/2)$, $\beta_0^2=r_g/r_0$, in accordance with the criterion of  $\sigma=\sigma'=1$. The reduction of the angular speed makes a margin to extend the period in $S'$ to its value in $S$. We think that the SR Thomas precession must take place and can be observed in a planetary motion.

The Thomas precession is loosely termed ``the orbital plane rotation in the direction of motion''. However, such a terminology belongs to classical physics language, which in no way recognizes the Minkowski space concepts including Lorentz transforms $S'\rightleftharpoons S$ and rescaling of physical units. 

\subsection{About relativistic terms}

Many  disputes about relativity  (not to speak about GR) are counter-productive just because of a loose terminology, an abuse the SR concepts and definitions, finally, neglect of operational meaning of the Lorentz transformations. See some examples below.

{\em Relativity}. One should bear in mind that the SR concept of ``relativity'' of motion cannot be ``refuted'' because it stands for the classical (Galilean) relativity, on which the SR is based (the postulate 1). The SR Theory cannot be refuted because it is based on the postulate 1 and 2, the second postulate being a constancy of speed of light in all inertial coordinate systems. Both postulates are firmly proved in experiments. Why the GR is ``{\em General} Relativity'' is actually not clear, see \cite{Fock}.

{\em Event}. A non-trivial ``event'' includes two events constituting the 4-coordinate and 4-momentum vectors transformed by the Lorentz operator. A vector must have the origin and the tip, the two events, which are: 

   1. synchronization of clocks in $S'$ and $S$ at $t'=t=0$, $x'=x=0$; 

   2. light signal exchange between $S'$ and $S$ observers. The concrete procedure depends on  an objective of observation (type of intended event).

{\em Coordinate time}. There is no such a thing as an observer's wristwatch recording the coordinate time $t$: the time is measured with the use of records of $S'$-clock at $x'$ and $S$-clock at $x$. In ``the pure Lorentz boost'', $y$ and $z$ coordinates are not involved. The time dilation and the length contraction take place only in the direction of motion. A motional (angular or linear) advance is an ``inverted'' length contraction.

{\em Clock paradox}. The problem with the paradox typically lurks in the ill-posed formulation and a potential abuse of the coordinate time concept. The matter is that a rest observer can work by herself  ``observing'' events from records of proper time intervals. Alternatively, she can be in couple with the inertial observer to collect data for measuring the coordinate time. The quantity to be compared is the Lorentz invariant one such as the proper time interval, or the phase of periodic events that is,  a number of clock ticks $N$ detected by a wristwatch of each observer. A different approach to the paradox formulation is a collection of periodic light pulses in the Doppler operational mode \cite{VanTwin}. The clock paradox can be particularly recognized in the Thomas precession problem.

\medskip

Further we continue to analyze the perihelion advance problem. So far, we failed in our efforts to realize the physical meaning of the GR-term at a deeper than algebraic equation level. We are looking with hope to finding answers in original works by Einstein and prominent GR experts. 
After that, we shall return to the question whether the Thomas precession can be seen  in the sky as a part or the whole relativistic gravitational effect.


\section{The GR-term with Einstein}

\subsection{The year 1915}

Einstein starts with the equation ({\ref{Ein1}), \cite{Einst1} (1915), which is derived in the scale of the proper time $\tau$. Still it is not clear why $\tau$ but not $t$. He treats the solution in terms of $e$-representation; hence, the equation is given by 


\begin{equation}
(du/d\theta)^2=    -(1- e^2)/p^2    +   (2/p)  u   -   u^2  +  
 \fbox{  \parbox{1cm}{$  2r_g u^3 $ } }  
\label{Ein2}
\end{equation}
where $u=1/r$, $e$ and $p$ are classical ellipse parameters. An elliptic solution has the form $u_{1,2}=u_0(1\pm e)$. The equation has three roots, $u_1$,  $u_2$,  $u_3$, where  $r_1$ and  $r_2$ are perihelion and apohelion points while $r_3$ has no classical analogy; its origin is due purely to the GR term, see the root $r^*$ in Fig. \ref{3roots}.


The Einstein's plan to solve the perihelion advance problem was straightforward : to integrate the arc path from perihelion to apohelion to see how the result differs from $\pi$. Many years later, the plan was reproduced by Moller  \cite{Moller}, (1972) with somehow different physical concepts. He came to the same integral, as Einstein did, but in the scale of coordinate time $t$. Anyway, the solution is
\begin{equation}
\theta= \int_{\tilde{u_1}}^{\tilde{u_2}} { du \left[(e^2 -1)/p^2 + (2/p) u -  2r_g u^3 \right]^ {-1/2} }
\label{Ein3}
\end{equation}
It should be noted that the Moller's derivation is presented in more details. A reader is recommended to read both works in parallel (Einstein's in German and Moller's in Englsh). Notice a missed factor $1/2$ in (16) in \cite{Einst1}.

Sadly, the Einstein's and then Moller's plan was not accompanied with discussions of the role of the GR framework and the corresponding initial conditions. As shown above, without comparing circular motion conditions in classical and relativistic variants (say, by $\sigma$ criterion), the problem becomes ill-posed, and a conclusive result cannot be achieved. The same remarks can be addressed to the Synge's work \cite{Synge}. Our critical argument on the methodology of the GR effect based on understanding of the fact that the relativistic angular advance (in both SR and GR methodologies) has no analogy in classical physics. The relativistic criterion of observability and treatment of the effect should be described in clear GR terms. Let us see how the plan proceeds. 

To avoid  complications with  the exact solution  (\ref{Ein3}), Einstein considers, as an approximation, the following integral
\begin{equation}
\theta= \int_{\tilde{u_1}}^{\tilde{u_2}} { du \left[-2r_g/r_0(u-{\tilde u_1})(u-{\tilde u_2})(u-{\tilde u_3}) \right]^ {-1/2} }
\label{Ein3sol}
\end{equation}
where ${\tilde{u_1}}$ and ${\tilde{u_2}}$ are the exact roots for (\ref{Ein1}). Then he decides to modify the polynomial of third degree in order to use the exact solution of the classical equation $u_1$ and $u_2$ (\ref{3a1}) assuming that the equalities ${\tilde{u_1}}=u_1$, ${\tilde{u_2}}=u_2$ are valid to the precision of smallness of the GR-term. In other words, Einstein (and later Moller) makes the assumption that the main roots $u_1$ and $u_2$ are not affected by the presence of the third root $u^*=1/r^*$ so that $u^* >>  (u_1 + u_2)$ to justify the assumption $(u^* + u_1 + u_2)= 1/(2r_g/r_0)$ for however small $e$. This makes the third root a known function of the main roots $u^*=f(u_1, u_2)$ so that the multiplier $\nu=(1-3r_g/p)$ appears in the argument of periodic function 
\begin{equation}
(1/\nu^2)(du/d\theta)^2=    (e^2 -1)/p^2    +   (2/p)  u   -   u^2  +  
 \fbox{  \parbox{1cm}{$  2r_g u^3 $ } }  
\label{Ein2rot1}
\end{equation}
the solution of which describes the elliptic motion with the argument ($\nu\theta$):
\begin{equation}
u=1/p+(e/p) cos{(\nu\theta)}
\label{Ein2rot2}
\end{equation}

We argue, however, that the Einstein's assumption is physically and logically inconsistent: if the third root $u^*$ (which is closely related to the GR term) does not affect the main roots $u_1$ and $u_2$, then it cannot be presented by a certain function of them. In fact, we know that the shift of main roots is about $3r_g$ what is comparable with the GR-term and the 3d root.

We state that the Einstein's wrong assumption and rude approximations lead to the false prediction of the perihelion advance $3r_g/r_0$. It can be verified by a direct comparison of the original algebraic expression with the approximate one obtained by Einstein  \cite{Einst1} (1915)
\begin{equation}
-2r_g \left[(u- u_1)(u- u_2) \right]^{-1/2}=(1+2r_g/p)(1+r_g u))\left[(u- u_1)(u- u_2) \right]^{-1/2}
\label{check1}
\end{equation}

Squaring and inverting both sides of (\ref{check1}) leads to the contradicting expression
\begin{equation}
0\equiv -2r_g/p
\label{check11}
\end{equation}
Our argument that the Einstein's  prediction is not valid can be also verified by a direct substitution of (\ref{Ein2rot2}) into  (\ref{Ein2rot1}). Having this done, one finds an exact cancellation of all terms on both sides of the equation but the GR term on the right:
\begin{equation}
0\equiv\fbox{  \parbox{1.8cm}{$ 3(r_g/r_0) u^2  $ } } 
\label{Ein8}
\end{equation}
Thus, the claimed formula for the perihelion advance $3r_g/r_0$ is a result of inappropriate mathematical assumption in the equation solution, therefore, the prediction is not valid.

\subsection{The year 1942}

Probably, Einstein did not feel a satisfaction with his account for the GR perihelion advance. In the following years he never returned to the above derivation in his publications and never suggested some modification of it. The paper  \cite{Einst1} (1915) was not translated into English, and still it remains hardly accessible. About a quarter of century later, the monograph on Relativity Theory by Bergman \cite{Bergmann}  (1942) was published where Einstein in the foreword acknowledged his personal role in its composing. There, for the first time another derivation of the effect is presented. It starts with the second order differential equation
 \begin{equation}
du^2/d\theta^2 = 1/p - u +   \fbox{  \parbox{1.1cm}{$ 3r_g u^2  $ } } 
\label{Ein4}
\end{equation}
where the GR term carries the coefficient responsible for the main root shift and the predicted effect. Instead of integration of the equation, the  author suggests that the  Fourier's expansion of the solution be analyzed in view of a small parameter $\lambda=3r_g/r_0$ (in his denotations):
\begin{equation}
u=\alpha_0 + \alpha_1 \cos{(\rho\theta)} + \lambda \sum_\nu {\nu^2\beta_\nu \cos{(\nu\rho\theta)}}
\label{Ein5}
\end{equation}
starting with  the first approximation
\begin{equation}
u=\alpha(1-e\cos)\theta
\label{Ein6}
\end{equation}
After making some more (not clear) approximations, the author comes to the solution obtained before (\ref{Ein2rot2}), which we have already proved to be wrong.

\subsection{Fock's work, 1955}

About 13 years after the Bergmann's work, the monograph on ``The Theory of Space, Time, and Gravitation'' by Russian physicist  V. Fock was published with a critical attitude towards the GR theory \cite{Fock}, (1955). There Fock suggests the concept of harmonic coordinate system, the idea of which reflects d'Alamberian wave conditions at infinity. Thereafter, the Schwarzschild metric undergoes some modification with the appearance of terms of higher (up to fourth) order in the equation (\ref{Ein1}). While considering the Mercury perihelion advance, Fock follows the Einstein's methodology of relativistic reduction of frequency in periodic motion $u=(1/p)(1-e\cos{(\nu\theta)}$ with $\nu=(1-3r_g/r_0)$  to provide the perihelion advance. His analysis of the problem in the harmonic coordinate system shows that a combination of quantities in the right side of the equation can match the multiplier  $\nu$ needed for the relativistic rotation of the orbital plane. However, the terms of 3d and 4d order must be neglected (the Einstein's GR-term included). It means that the GR term plays no role in the effect formation. However, the introduction of ``harmonic conditions'' can make a difference.

\subsection{The GR problem of the perihelion advance: summing up}

We state that the GR prediction is in a serious doubt. First of all, the problem  is ill-posed: its formulation does not allow an observer to differ between a small GR effect from similar classical effects.  Conditions, under which the  GR effect acquires relativistic properties consistently with the GR framework, are not specified (for example, the role of $\tau$ versus $t$ scaling). There is no answer to the question whether the kinematical effect of the Thomas precession contributes to the GR effect in question. Finally, technical problems in the derivation of the GR effect should be noted. Namely, the constraints imposed by the $e$-classification of the orbit make a physical  analysis of the problem ambiguous.  The actual derivation of the solution is flawed.  Overall, the claimed prediction is proved physically invalid.

Our expectations to clarify the problem formulation from the original  Einstein's work and works by reputed GR experts over decades after the Einstein's work did not come true.  In literature, authors present derivations of the effect with a variety of different techniques, but always with the same result.  At the same time, they do not attempt to analyze factors of GR spacetime curvature, which would create the effect. No one wanted to check the solution by numerical computations or by its substitution to the original equation. The hard questions remain if the effect can survive without the GR-term (the Fock's work), why the effect in most of the works is derived in the scale of proper time  $\tau$ rather than the coordinate time $t$ (the Moller's work), finally, what role, if any, plays the GR-term in the GR testable effects with light (to examine immediately). 

The above statements, if true, are too serious and raise questions about validity of the GR physical foundations. Though it is out of the scope of the present work, some issues of GR foundations will be briefly discussed in the remaining part of the work. In particular, we want to investigate a role of the GR framework in providing a unified approach to problems involving particles and photons.


\section{GR tests with light} 

\subsection{The GR-term as the cause of the bending of light} 

Previously, it was shown how the GR equations of motion of the particle and the photon are derived from the  Hamilton's  principle of variation applied to the Schwarzschild metric. The two conserved quantities are found: $\epsilon$, associated with the energy (\ref{conser1}), and $l_0$ -- the angular momentum  (\ref{conser2}). The remarkable (and strange) fact is that those quantities are the same for the particle and the photon 
$$ \epsilon=(dt/d\tau)/\Gamma_r^2, \  \  \  l_0=r^2 (d\theta/d\tau) $$
 This is strange because the angular momentum is given per unit mass, and the question arises, what mass the photon has. 
The equation  (\ref{pho}) describing the bending of light in the Schwarzschild field is
$$(du/d\theta)^2=\epsilon_0^2/l_0^2  - u^2  +   
\fbox{  \parbox{1.1cm}{$  2 r_g u^3 $ }  }$$
Now a reader should be especially attentive. Unlike in the case of particle, there is no linear term there related to the gravitational potential. Hence, the GR-term solely is supposed to be the cause of the bending of light. Indeed, putting the term to zero results in the solution describing a straight line trajectory of the photon similarly to the classical photon motion in ``empty'' Euclidean space. Further we shall see grave consequences of the lost linear term.

\subsection{Einstein's evaluation of the light bending, years 1911, 1916, 1947}

The Einstein's first assessment of the bending of light grazing the Sun is given in  \cite{EinBend}, (1911), $\Delta \theta=2 r_g/R$, where $R$ is the Sun's radius.  Though Einstein refers to the Huyghen's principle he actually uses the idea of light refraction with the index $n_r=1/\beta$ ((the Snell's law)) to integrate the angle along the light path. This means that the speed of light depends on  the potential $\beta(r)=\beta_0 (1-r_g/r)$. The result was widely criticized at that time because it was exactly the same as obtained in the Newtonian model years before. 

In the second paper devoted to the GR foundations \cite{Ein1916},(1916) Einstein corrected the calculation. Again he refers  to the Huyghen principle but makes calculations of the same integral with the redefined index of refraction  $n_r\approx(-g_{44}/g_{22})^{1/2} \approx (1+2r_g/r)$. This should double the effect making the full angle of deflection (between two asymptotes) 4$r_g/R$, about 1.7$''$. 

As it seen, both earlier and this evaluation has nothing to do with the above discussed GR framework.  Einstein, when working on the perihelion advance problem in 1915, had to be aware of the space-time symmetries, from which the equations of motion (\ref{tau}) and the trajectory for both {\em the particle and the photon} were deduced (\ref{Ein1}), (\ref{Ein2}). However, his derivation of the light deflection in 1916 was made in the {\em ad hoc} approach, outside the above GR framework. The effect about 1.7$''$ of deflection was confirmed in a series of observations of total eclipses of the Sun's, but this confirmation was made against the logic of GR physical foundations rather than in accordance with that. This circumstance remained largely unnoticed (or thought excusable?) in the GR history.

Undoubtedly, Einstein realized the methodological inconsistency of his predictions and the fact that the derivation of the bending of light had to be reexamined and formulated in the unified GR framework given by (\ref{conser1}), (\ref{conser2}), (\ref{tau}),  (\ref{Ein1}),  (\ref{pho}). About fifteen years passed before it was made. In 1942, Bergmann (with Einstein) \cite{Bergmann} abandoned the idea of refracting deflection and finally turned to the equation (\ref{pho}) in order to evaluate the bending of light
\begin{equation}
(du/d\theta)^2=\epsilon_0^2/l_0^2  - u^2  +   
\fbox{  \parbox{1.1cm}{$  2 r_g u^3 $ }  } 
\label{PhCheck}
\end{equation}
The approach used this time is consistent with that for the perihelion advance problem. But now (recall our warning) one encounters a severe problem of absence of the potential term in the photon trajectory equation. The only term left to cause the effect is the GR-term. The latter, as we found, plays a controversial role in the perihelion advance problem. In particular, the result of Fock's work, \cite{Fock} (1955) is  remarkable: the perihelion advance is explained due to the Schwarzschild metric adjustment to the harmonic coordinate system, while the GR-term is ignored. As concerns the role of potential energy term in the bending of light, see also the textbook \cite{Misner} (1970):

``Relativistic effects [in GR] on light and radio-wave propagation are governed entirely by the Newton potential $U$ and the PPN parameter $\gamma$''.

The matter is that the authors \cite{Misner} were not inclined to investigate the GR framework; they evaluated all the GR classical effects in the PPM formalism. We do not know how the potential term was reborn there.

In 1942, nothing could be done but take the same plan as in the perihelion advance problem: to find corrections to the exact unperturbed solution (a straight line for $r_g=0$) due to the ``perturbation effect'' caused by the small GR-term. Interpolations are used based on the assumption that the exact roots of (\ref{PhCheck}) are not influenced by the small term. 

The author \cite{Bergmann} arrives at a ``correct'' result $\Delta\theta=4r_g/R$= 1.7$''$ with the use of procedure, which is hard to follow. It should be noted that the solution was ``reproduced'' later, for example Moller \cite{Moller} (1972). In \cite{Wald}, also  \cite{Taylor}, the GR concept of the effective potential $V_{ef}$ is used to get the same equation. The PPN formalism was used in several works, \cite{Misner}, \cite{Weinberg}, and other works. They are not discussed here for the reasons explained earlier.

In order to see the inherent inconsistency of the Bergmann (with Einstein) solution of the light bending problem, one has to unfold the algebraic procedure to the point of checking the result by substituting it back into the original equation (as we did in the case of Mercury's perihelion advance). We found that the claimed solution is not only technically but fundamentally wrong because of the loss of the potential energy term. We need now to discuss the Fock's work \cite{Fock} (1955), which brings a lot of clarity in this problem.

\subsection{Fock's work, 1955}

The Schwarzschild metric in Fock's harmonic coordinates \cite{Fock} is given by
\begin{equation}
d\tau^2=\left(\frac{r-r_g}{r+r_g}\right)dt^2-\left(\frac{r+r_g}{r-r_g}\right)dr^2-(r+r_g)^2d\theta^2
\label{FockShwa}
\end{equation}
The photon equation follows from (\ref{FockShwa}) after setting up the boundary condition and neglecting terms of orders higher than $\sim u^2$
\begin{equation}
(du/d\theta)^2=1/R^2 -u^2 +4 r_g u/R^2
\label{FockPhoEq}
\end{equation}
Compare it with  (\ref{PhCheck}) (Einstein 1947, Moller 1972, Wald 1984)  
$$(du/d\theta)^2=1/R^2  - u^2  +   
\fbox{  \parbox{1.1cm}{$  2 r_g u^3 $ }  } $$

Here $R$ is the impact parameter (the Sun's radius in eclipse observations). In the Fock's equation, this is a liner (potential energy) term that makes the light bending effect. The solution is
\begin{equation}
R u=2 r_g/R +\cos\theta
\label{FockPhoSo}
\end{equation}
It describes a hyperbolic trajectory. The full bending angle (between two asymptotes) is 4$r_g/R$=1.7$''$. 

Fock's comments are, as follows. The effect can be treated in terms of refraction  in the gravitational field when the latter is considered the optical active medium with a continuous change of the index of refraction $n_r=1/\beta(r)$. This suggests that the speed of light decreases with the potential depth. If the effect is treated in terms of gravitating photon in Newtonian Physics (the speed dependence on the potential only), then the effect would be half the full one (the value obtained by Einstein in 1911). The photon motion equation derived from the Schwarzschild metric {\em in the harmonic coordinate system} gives the correct effect due to the potential energy term leading to $n_r=\Gamma_r^2$ that is, $\beta(r)=1/\Gamma^2=(1-2r_g/r)$. However, usage of the Schwarzschild metric in ``the standard coordinate system'' (\ref{Schwa}) results in the motion equation  (\ref{pho}) with the potential energy term lost. 

It looks like the Fock's harmonic conditions make a difference. In connection with the choice of coordinate conditions in GR, it is usually emphasized the fact that the physical world does not depend on our choice of coordinate system (the general covariance). This would be a simplification of the problem. In GR, not just a coordinate system should be fixed. In order to uniquely solve the Einstein's field equations, four differential equations are provided that the metric tensor must satisfy. Therefore, the coordinate conditions have to impose physically motivated constraints on solutions. 

The Fock's harmonic coordinate system is not original; it is also known from the earlier suggested de Donder's gauge: the 4-coordinates, which are Lorentz invariant solutions to the d'Alamber's equation. In \cite{Fock}, motivations for choosing the harmonic conditions are thoroughly discussed, in particular, in connection with the conditions at infinity.

\subsection{The Light and the Particle in GR}

\subsubsection{Brief conceptual review of the GR tests with light}

In the Schwarzschild field, there is a symmetry between the photon and the particle. Expressions for their total energy and angular momentum are identical. The only difference is in that one has to put $d\tau=0$ for the photon. The photon can be treated as the particle in a zero mass limit still not loosing a property of attraction by the gravitational source (because the energy is a source of gravitational field in GR). In particular, the photon can be launched into an orbit in a particle manner. 

However, the GR photon concept $d\tau=0$ does not define the photon wave properties such as frequency and wavelength. A connection of those properties with a photon source is not recognized either. Moreover, the de Broglie wave concept in the modern Special Relativity theory is not respected because the Special Relativity theory is declared incompatible with the gravity phenomenon, paradoxically, for an alleged failure to describe physical properties of the light, or photon, \cite{Misner}. At the same time, we just witnessed the GR failure to explain observations of the the light bending effect, where wave properties are most likely involved. What about the red-shift effect?

\subsubsection{The red shift}

Consider the photon emitted from the Sun's surface and observed on Earth (the GR red-shift test). The Einstein's explanation of the observation \cite{Ein1916} (1916) is based on the comparison of frequencies of the standard clock and the photon. Let the standard clock have the frequency 
$f_1 (cl)$, say, on the Sun's surface $r_1$ that is $f_1 (cl)=f_0 (1-r_g/r_1)$. Then he notes that the photon of the same frequency $f_1 (ph)$ travels from $r_1$ to $r_2$, (say, the Earth) and is detected there with the same frequency $f_1 (ph)=f_2 (ph)$, while the clock at $r_2$ runs faster, namely, $f_2 (cl)=f_0 (cl)(1-r_g/r_2 $, $f_0$ is the clock rate at infinity. However, he could give  another scenario of the event development with the same assurance, as next. A standard photon source is not influenced by the gravitational field. This is the frequency of the photon in flight, which depends on the potential. This would also explain why the Earth's observer detects red-shift in a star light.

The GR expertise ``explanatory'' comments  on this issue \cite{Okun1}, \cite{Weinberg} do not and cannot bring clarity because gravitational properties of the photons and the photon emitters are not defined in GR. A relationship between the photon's gravitational characterisics such as the speed $\beta(r)$, the frequency $f(r)$, and the total energy  $\epsilon(r)=(dt/d\tau)(1-2 r_g/r)$ should be brought into the GR framework. 

Let us sum up the bending light and the red-shift issues. In the Bergmann-Einstein's explanation  \cite{Bergmann} (1942) of the light bending effect, the photon is treated as a particle having no wave properties, consistently with the GR framework (\ref{Schwa}), (\ref{conser1}), (\ref{conser2}), (\ref{Ein1}), (\ref{pho}). The Focks' result (under harmonic coordinate conditions) also was obtained without any additional postulates. The qustion arises why the predictions are different.  

The author \cite{Okun1} seems to be aware of the controversy and suggests to modify the GR foundations, namely, to add a postulate connecting gravitational properties of a photon source and the photon. Specifically, the frequency of the emitter of the standard photon should be a function of the gravitational potential $f(r)=f_0 (1-r_g/r)$, while the frequency of the photon in flight does not change. 


\subsubsection{The particle in free radial fall}

As previously shown, the prediction of GR perihelion advance caused by the GR-term is not valid. We state that the particle orbit equation, though being consistent with the GR framework, is physically wrong. There is another GR prediction rarely discussed in literature. It concerns a particle radial motion in the gravitational field. The GR theory predicts that the particle approaching the center in free fall {\em decelerates}, consistently with the GR framework. This effect is, in principle, testable, for example, in observations of cosmic high-energy particle coming onto Earth.   

Specifically,  \cite{Misner}, also  \cite{Taylor}, \cite{Okun2}, the speed of a particle in the radial fall is  
\begin{equation}
\beta(r)=(1-2r_g/r)[1-(1-2r_g/r)/\gamma_0^2]^{1/2}
\label{RadFal}
\end{equation}
Here, $\gamma= E_0/m_0>1$ is the initial total energy at infinity. A free fall from rest corresponds to $\gamma_0=1$. The formula is given in a coordinate system of the observer at infinity at rest with respect to the gravitational center. Therefore, her wristwatch time is the coordinate time $t$.

The formula shows \cite{Okun2} that the particle sent from infinity to the gravity center begins to accelerate, then at some point  starts decelerating and eventually stops at $r_d=2r_g$. The higher initial kinetic energy, the farther the point of $r_d$ from the center. For $\gamma_0 \ge\sqrt{3/2}$, the particle will never accelerate in a gravitational field. The gravitational force exerted on the particle becomes  repulsive in the entire space.  We think, however, that there is no real physics behind this prediction \cite{VanGrav}. 

One should notice that in GR problems the quantity $\epsilon$ plays a special role in the formulation of initial conditions. Next, a controversy with GR concept of energy $\epsilon=(dt/d\tau)/\Gamma_r^2$ is discussed.

\subsection{The GR energy and the initial value}

Recall the GR scheme: from the Schwarzschild metric (\ref{Schwa})

$$ d\tau^2=dt^2/\Gamma_r^2-\Gamma_r^2 dr^2-r^2 d\theta^2$$ 
with 
$$\Gamma_r^2=1/(1-2r_g/r)\ge 1$$ 
 get (\ref{conser1}), (\ref{conser2}) 
$$  \epsilon=(dt/d\tau)/\Gamma_r^2; \  \  \  \    l_0=r^2 (d\theta/d\tau) $$ 
and the orbit  equation (\ref{Ein1}) with the GR-term
$$(du/d\theta)^2=-(1-\epsilon_0^2)/l_0^2 + 2(r_g/l_0^2) u - u^2  +   
\fbox{  \parbox{1.1cm}{ $ 2 r_g u^3 $ }  } $$
Typically, $ \epsilon $ is absorbed in the orbit equation without need to specify the quantity $(dt/d\tau)$.
But often one needs to express the initial values $l_0^2$ and 
$\epsilon= \left[(dt/d\tau)(1-2r_g/r)\right]_0 $ in terms of $r_0$, $\theta_0$ and the squared speed $\beta_0^2 = (\beta_r^2 + \beta_\theta^2)_0$ in an arbitrary geometrcal configuration.
In GR text-books, for example \cite{Taylor} and elsewhere, there are instructions on how to do this. 
It is suggested to introduce ``the shell observer'' who accounts for the proper time $\tau_{sh}$ with respect to the variable proper time $\tau$, also, keeps connections with the coordinate time $t$, as next    
\begin{equation}
(dt/d\tau)=(dt/d\tau_{sh}) (d\tau_{sh}/d\tau)
\label{ic2}
\end{equation}
where
\begin{equation}
(dt/d\tau_{sh})= (1-2r_g/r)^{-1/2};\  \   (d\tau_{sh}/d\tau)=(1-\beta_r^2-\beta_\theta^2)^{-1/2}
\label{iic2}
\end{equation}
Hence
 \begin{equation}
 (dt/d\tau)^2=(1-2r_g/r)/(1-\beta_r^2-\beta_\theta^2)
\label{ic3}
\end{equation}
where $\beta_\theta^2=l_0^2/r^2$.  

As a by-product of the above ``instruction'', we have a new expression
 \begin{equation}
\epsilon_0^2=\left[(1-2r_g/r)/(1-\beta_r^2-\l_0^2/r^2)\right]_{i.c.}=
(1-2r_g/r)/(1-\beta_r^2-l_0^2/r^2)
\label{ic4}
\end{equation}
From (\ref{ic4}:
\begin{equation}
 \beta_r^2= (\epsilon_0^2-1)/\epsilon_0^2 + 2r_g/\epsilon_0^2 r - l_0^2/r^2 +   
\fbox{  \parbox{.4cm}{ $  $ }  } 
\label{ic5}
\end{equation}
But where is the GR-term? It is gone. Recall, the GR-term appeared in the equation of motion from the Schwarzschild metric, and ``the instruction'' is supposed to be consistent with the GR framework. But it is not. If ``the instruction'' has a physical meaning and removes the GR-term from the box, it explains why Einstein and other authors suffered so much with the GR term. It looks that the GR field equations and the following from them Schwarzshild metric have a serious initial value problem.

\section{Different Theory?}

We propose another (relativistic gravitational) theory \cite{VanGrav}. In the referred work, the principles of relativistic dynamics the theory based on are given and illustrated in the example of free radial fall in the spherical symmetric field. The orbital motion problem is described in the present work, as we planned. 

In brief, the principles are, as follows. A new concept of the relativistic proper mass $m(r)$ depending on field strength is introduced. From the Lagrangian problem formulation, it follows $m(r)=m_0/ \gamma_r$  where $m(r)\to m_0$ as $r\to\infty$, with $\gamma_r= \exp({r_g/r})$. The revision of the proper mass concept is motivated by several reasons, one of them, a necessity to introduce the 4-momentum vector $P^\mu$ in the form complementary to the 4-coordinate vector $X^\mu$. The temporal component in  $X^\mu$ is the proper time depending on the gravitational potential $\tau=\tau(r_g/r)$. Therefore, the temporal component $m$ in $P^\mu$ should be $m=m(r_g/r$. This explains the gravitational time dilation. 

Thus, the gravitational dynamics is formulated in the Minkowsli space in the presence of gravitational sources. In polar coordinates, the 4-coordinate interval and the 4-momentum vectors are $dX^\mu (r)=\gamma d\tau (r)\  (1,\ \beta_r,\ \beta_\theta)$ and  $P^\mu (r)= \gamma m(r)\ (1,\ \beta_r,\ \beta_\theta)$ , where 3-velocity components and the Lorentz factor are functions of $r$ and $\theta$, $c_0=1$. The Minkowski 4-force $K^\mu= dP^\mu/d\tau$ acts on the test particle, and it naturally has the tangential component (with respect to the world-line $s$) and the orthogonal one, while $s$ is a function of 4-position. 

There are two conservation laws, -- for total energy $\epsilon_0$, and the angular momentum $L_0$  given below for initial conditions $r(r)=r_0$, $\theta=0$, $\beta_r=0$, $\beta_\theta=\beta_0$

The total energy and the angular momentum are
\begin{equation}
\epsilon_0=\gamma_0 {\gamma_r}_0=\gamma\gamma_r
\label{a1}
\end{equation}
\begin{equation}
L_0=\gamma_0 {\gamma_r}_0 r_0 \beta_0=\gamma\gamma_r r \beta_{\theta}
\label{a2}
\end{equation}
Instead of (\ref{a2}), it is convenient to use a conserved quantity $l_0=\epsilon_0/L_0$:
\begin{equation}
l_0=r \beta_{\theta}
\label{a3}
\end{equation}
Here, a squared inverted Lorentz factor is $1/\gamma^2=(1-\beta_r^2-\beta_\theta^2)$,\ $\beta_r=dr/dt$,\ $\beta_\theta=r (d\theta/dt)$, and we are going to use, as usual, the formula  $\beta_r=(dr/d\theta )(d\theta/dt)$, and $\beta_{\theta}^2= l_0^2/r^2$. After introducing a variable $\xi=r_0/r$, we arrive to the exact relativistic equation of orbital motion of confined particle. The equation has the Newtonian limit, and it is valid for a however strong field (by the criterion $r_g/r$).
\begin{equation}
\left(\frac {d\xi}{d\theta}\right)^2=\left(\frac{1}{\beta_0^2} -\xi^2 \right)  - \left(\frac{1}{\gamma_0^2\beta_0^2 } \right)  \exp{\left(\frac{2r_g}{r_0} (1-\xi) \right)}
\label{a4}
\end{equation}
The Newtonian limit (weak field conditions) is given by a linear approximation of the exponential function
\begin{equation}
(d\xi/d\theta)^2=(1-2\sigma_r) + 2\sigma_r \xi - \xi^2
\label{a5}
\end{equation}
where $\sigma_r =r_g/r_0\gamma_0^2\beta_0^2$ is the $\sigma$ criterion in the relativistic case. 

Here, we recognize the Thomas precession proportional to $ r_g/2r_0$ in the gravitational field due to the point mass $M$ creating the potential energy field $r_g/r$. In the strong field, the potential is
\begin{equation}
V(r)=-\left(1-\exp(-r_g/r)\right)
\label{a6}
\end{equation}
Therefore, in the Newtonian limit, the relativistic Thomas precession is caused by the SR Kinematics. Next order corrections will be due to the deviation of the gravitational potential from the classical law $V(r)\sim ~1/r$. Definitely, we do not confirm the GR precession proportional to $3r_g/r_0$.

The theory predict a free radial fall without the deceleration
\begin{equation}
\beta(r)=\left[1-(1/\gamma_0^2)\exp(-2r_g/r)\right]^{1/2}
\label{a7}
\end{equation}

Few more words about the theory. It requires a revision of the conventional relativistic concept of mass: the constancy of the proper mass in any field of forces. The revision results in an elimination of divergence of gravitational potential at $r\to 0$.  It is shown that the introduction of the concept of variable proper mass also resolves the long standing problem of divergence in relativistic electrodynamics. For years, the problem has been tackled by the artificial procedure of mass/charge renormalization (by means of subtractions of infinite numbers). 

The theory adopts a field concept as an optical active medium for propagation of electromagnetic waves and material de Broglie waves. In this way, it has quantum connections needed for further development of relativistic gravitational field theory. The theory denies the GR ``Black Hole'' concept of gravitational collapse of matter and light. It explains all GR classical tests and has new practically verifiable predictions.


\section{Conclusion}

We  raised questions, which are routine for any physical (small or great) finding:
what and how it is observed, what and how it is predicted, finally, how both sides are matched by a statistical method of empirical data treatment. The results of our investigation is summarized below. 
\medskip

{\em Astronomical observations of the GR perihelion advance}.
\medskip

A direct observation of the effect is difficult because of its smallness on the background of classical perturbation and precession of the Mercury's orbit. The effect can be, in principle, assessed from the analysis of astronomical observations and the correspondingly adjusted ephemerides with the use of statistical methods.  The ephemerides production is a multi-aspect complex work, and there had to be a spacial research program devoted to the Einstein's prediction of fundamental importance. In reality, there was no such a program. There are many published claims and statements, but there is no work which would present results of the whole problem study.

Historically, sporadic initiative studies of individual researchers or groups were conducted. The goal mostly pursued  was ``to fit'' the Einstein's effect value to the ``anomaly gap'' as close as possible. No rigorous methods of empirical data treatment from statistical theories were used by researchers for the effect verification. The implementation of the PPN formalism in the ephemerides systems made the situation worse because of uncontrolled unphysical shift in calculational results. 

 Sadly, the opinion on the firm confirmation of the GR perihelion advance spread among the astronomical and physical communities. We state, however, that the real precision of current ephemerides is unknown, and our knowledge about the effect verification in the astronomical observations remains, roughly, at the Le Verrier's level.

\medskip

{\em  GR prediction of the effect}. 

 The GR  description of the perihelion advance is given in such a form that the  effect cannot be discriminated from the huge fluctuating background of classically induced precession. Consequently, astronomers had to identify the effect as a difference of big fluctuating numbers in observed and calculated ephemerides values. As emphasized, they tried ``to fit'' theoretical ephemerides to the exactly predicted number. This methodology and the corresponding results cannot be termed an observational test.

 Rigor of the effect prediction is in a serious doubt. The Einstein's ``approximate'' solution, when put back into the original equation, does not fit the equation to the precision better than the effect value. The GR-term, which is thought to be the cause of both the perihelion advance effect and the light bending effect, has no physical sense, as shown by Fock and in our work. The Fock's work also shows that the equation of light propagation in a vicinity of massive object must have the linear (potential) term instead of the GR-term. The latter is of the next degree of smallness and should be neglected. 

Controversies in the GR theory arise in connection with gravitational properties of both the particle and the photon. The GR prediction of the particle deceleration in a radial free fall has no physical sense and most likely is wrong. The prediction of the bending of light is not valid for the same reason as in the perihelion advance case, while the predicted red-shift is inconsistent with the GR framework. These and other arguments raise the question about sufficiency and completeness of the GR physical foundations.

Overall, we conclude that the claimed confirmation of the GR prediction of the relativistic perihelion advance is neither theoretically nor empirically substantiated.

\appendix

\section{Rigorous Statistical Approach}

\subsection{Definitions and general formulation}

Information on the general statistical concepts and the corresponding method of measured data treatment are available in literature, for example \cite{Statistics} and elsewhere. A presented statistical method is applicable in general case of any active experiments, passive observations, and any kind of measurements, assessments of experiment informativeness, planing experiments, etc. Roughly, the problem is formulated as next. Suppose, some observable system is studied.  The empirical data include measured quantities and their statistical dispersions with possible correlations in the form of covariance matrices. One needs to assess system parameters in order to predict system characteristics, which are, in general, functions of spatial coordinates and time with constraints imposed on a set of parameters within the model.

\medskip

Definitions and denotations. 

\medskip

$X$ $=>$ $ X_i\ (i=1,\ 2...\ I)$ -- a set of model parameters (masses $M$, $m_i$ and others);  

$F(X)$ $=>$ $ F_k\ (k=1,\ 2...\ K)$ -- a set of observables, including ephemerides; they are functions of time $t$. 

$S(F|X)$ $=>$  $ S_{ki}= \left[  \partial{F_k}  /  \partial{X_i }   \right]_{X=X_0} $   -- sensitivity coefficients. They are obtained in numerical calculations.

$D(X)$, $D(F)$ $=>$ $ D_{rs} $ --  covariance matrices; they are quadratic symmetric positive-definite matrices that describe dispersions of parameters and their correlations before the experiment (prior information) and after (posterior information). They are obtained for the prior parameters $D(X_0)$, the measured quantities $D(\hat F)$ as well the corresponding posterior estimates  $D(X^*)$ and $D(F^*)$. A size of $D(X)$ is determined by a number of parameters $I$, and for $D(F)$ the number of measured quantities $K$. 
 
Matrix denotations here are similar to that in the linear algebra, where a row and a column are considered vectors what gives a conventionally defined operations of their dot-product and matrix multiplication. The  superscripts $<T>$ and $<-1>$ are used for transposed and inversed matrices. The method is designed to find the best estimates of model parameters' values and their dispersions and correlations arising from empirical data taken into account. 

The starting point is the introduction of the so-called likelihood function ${\cal P} (F, X)$, the criterion of matching measured results $\hat F$ to the corresponding calculational quantities $F(X)$. This is the normal distribution of the random quantity characterizing $|O(X)-C(X)|^2$ in terms of sensitivity coefficients $S(F|X)$ and the statistical weights (the inverse covariance matrices) for measured quantities  $D^{-1}(\hat F)$ as well as the prior data $D^{-1}(X_0)$. The initial dispersions and correlations are found from covariance matrices $D(\hat F)$ and $D(X_0)$. 

Thus, one needs to solve the extremum problem for the statistical distribution function
\begin{equation}
{\cal P}(F, X)=Const\ \exp{ \left[-(1/2)\Delta F^T D(\hat F)^{-1} \Delta F \right] } 
\label{P(F)}
\end{equation}
where 

$\Delta F =[\delta F- S(F|X) \delta X]$ is a random deviation of quantities $F(X)$ from observations, see next.  

$\delta F =(F(X_0)-\hat F)$ are differences between calculated and measured quantities (must be small deviations);

$\delta X = (X-X_0) $ are random deviations of parameters $X$ from their prior estimates (must be small deviations). 

Measured quantities are, in general, non-linear functions, or functionals (functions of functions) of parameters $F(X)$. To be consistent with the law of normal distribution, a statistical procedure of parameter evaluation should be conducted with a system of linearized functions 
\begin{equation}
F_k(X_i)=F_k(X_{i0})+  \left[ \partial{F_k}  /  \partial{X_i }   \right]_{X=X_{i0}} \delta (X_i-X_{i0})  
\label{Taylor1}
\end{equation}
or in matrix denotations
\begin{equation}
F(X)=F(X_0)+  S(F|X)_{X=X_{i0}} \delta X  
\label{Taylor2}
\end{equation}
This is how the sensitivity coefficients come out.

Prior data $X_0$ are assumed to obey a normal distribution law too:
\begin{equation}
{\cal P}(X)=Const\ \exp{ \left[-(1/2)\delta X^T D(X_0)^{-1} \delta X \right] } 
\label{P(X)}
\end{equation}
where $\delta X =(X-X_0)$. The criterion of  ``smallness'' means that deviations from ``true'' valued of the parameters due to uncertainties are small enough to validate a linear approximation of functionals $F(X)$ at the initial point $X=X_0$. If the point was appreciably shifted during corrections, sensitivity coefficients had to be recalculated at a new point. 

\subsection{Data statistical treatment in the Bayesian approach}

In the Bayesian approach to statistics, posterior and prior information is connected through conditional probabilities
${\cal P}(X |{\hat F)}{\cdot}{\cal P (F)}={\cal P}(F|{\hat X}){\cdot}{\cal P}( X)$. Hence, the posterior estimate is related to the distribution 
\begin{equation}
 {\cal P}(X | F)= Const\ \ {\cal P}(F| X){\cdot}{\cal P}( X)
\label{P(X|F1)}
\end{equation}
or in our case
\begin{equation}
{\cal P}(X|{\hat F})=Const\ \exp{ \left[-(1/2)\delta X^T D(X_0)^{-1} \delta X \right] } 
\label{P(X|F)}
\end{equation}

From the above formulas, the maximum likelihood principle leads to the posterior estimate of parameters
\begin{equation}
X^* =X_0+D(X_0)\left[S D(X_0)S^T+D({\hat F})\right]^{-1} ({\hat F}-F(X_0) 
\label{MeanP}
\end{equation}
The introduction of prior information into the assessment procedure gives an opportunity to conduct data treatment continually by portions as soon as new observational data become available. The prior data (mean parameters and covariances) are those assessed prior to new results come into the data base. With new results taken into account, the prior data become the posterior ones till next new results come, and so forth. Comparing the statistical standard error with the difference $(O-C)$ one can make conclusions on data consistency.

To evaluate informativeness and statistical significance of measurements, one needs to examine  the prior and posterior covariance matrices for sets of parameters as well as measured and predicted quantities. For parameters, it is 
\begin{equation}
{D(X^*)} =\left[ D(X_0)^{-1} +  S^T D({\hat F})^{-1}S \right]^{-1} 
\label{CovX}
\end{equation}
It can be given in the equivalent form that can be used to avoid an inversion of large matrices 
\begin{equation}
{D(X^*)} =  D(X_0)-  D(X_0) S^T \left[S D (X_0) S^T + D({\hat F}\right]^{-1} 
\label{CovP1}
\end{equation}
The assessments of new quantities (functionals) $Y$ at $X=X_0$ with sensitivities $S(Y|X)$ are similarly conducted.

To account for the rapid periodic and slow evolution of the system, one has to calculate sensitivity coefficients as functions of time. There is a problem related to the coordinate system transforms, which lead to the change of sensitivity coefficients. They should be recalculated in a new coordinate system and properly used in statistical evaluation procedures. It looks that our sketch suggests a lot of work one could try to avoid without a loss of already achieved apparently great values.

The last formulas are used in practical evaluations with the input data $X_0$, $D(X_0)$, $F(X_0)$, $\hat F(X)$, $D\hat F(X)$ and the output data $X^*$, $D(X^*)$, $F(X^*)$, $D(F^*)$. Notice that the covariance matrices are used to assess the posterior parameters while the latter are not needed to assess the covariance matrices. The sensitivity coefficients are used at all stages of data treatment. In the discussed problem, an analytical model is not available in practice; the coefficients should be found in numerical calculations $ S(F|X)= \Delta F/\Delta X)$. 

It is more convenient to work with relative quantities, namely, the input  $x=(X/X_0 - 1)$ and the output $f=(F/{\hat F} - 1)$, correspondingly. Then the sensitivity coefficients must be defined in the dimensionless form too $ S_{ki}= \left[ \frac  { \partial{F_k}}{F_k }  / \frac{ \partial{X_i }} { X_i }  \right]_{X=X_0} $, and the output data (adjusted parameters) are $x^*=(X^*/X_0-1)$. All covariance matrices will be related to the corresponding dimensionless quantities. 

It should be noted that the sensitivity coefficients can be geometrically visualized in terms of the linear (Euclidean) $N$D vector space with the corresponding linear algebra used above. Each parameter $P_i$ is associated with a unit vector $\vec n_i$ in a basis of linear independent vectors so that a function (or a functional) $F_k(P)$ is characterized by a vector $S_{ki}$ being a linear superposition of basis vectors. The covariance matrix describe a dispersion ellipsoid (standard errors and correlations) of elements in a subsets of vectors (measured or predicted quantities, parameters, and others). A correlation of two quantities $A$ and $B$ are characterized by a cosine of angle between vectors $\vec {A}$ and $\vec {B}$. Hence,  $A$ and $B$ are 100 $\%$ correlated, if collinear. A decrease of the ellipsoid volume (tnhe determinant of a covariance matrix) in the output of the experimental data treatment characterizes the precision improvement. In this way, the concepts of informativeness and information are rigorously defined.  
We would like to put an emphasis on the Bayesian approach to the statistical treatment of empirical data, as opposed to the the ``classical'' least square method. The latter reduces the problem formulation to the solution of $N$ algebraic equations $y_i=\sum_j {s_{ij} x_j}$ with $N$ unknown quantities (parameters) $x_n$ by inversion of the matrix of the system ${\hat a}=>\ {\hat a^{-1}}$ (the matrix of sensitivity coefficients). Often, one or several parameters happened to be ``ill determined'' because measured data $y$ are not sufficiently sensitive to them. This would result in the matrix determinant being close to zero, $det(\hat a)<<1$ (the  unfolding problem). The method fails.
 The use of prior information on the parameters in the first place is a physically natural method of statistical regularization of the solution (it makes the problem formulation ``well posed'').

\subsection{Role of constraints and initial conditions}

Let us discuss the role of constraints imposed on a system. As a simple example, consider a nuclear reactor model described by a set of nuclear constants $X_0$ having statistical dispersions $D(X_0)$  (in terms of a covariance matrix). Calculated characteristics $F(X_0)$ of a reactor are subject to reassessment with improved precision as a result of experiments. In the list of main characteristics is the neutron multiplication (effective) factor $K_{ef}$, calculated value of which depends on nuclear constants and parameters. In practice, the reactor is designed to make a neutron balance in a stationary state that is,  $K_{ef}=Const$. This requirement imposes a constraint on a variation of constants to be taken into account in the sensitivity coefficient calculations. Namely, a variable parameter $q(X)$ (usually, a control rod position in the reactor, a function of $X$) must be chosen to restore the condition the neutron balance when a single constant $X_i$ is varied:
 
\begin{equation} 
 0=\partial{k_{ef}}/\partial{X_i} + \left(\partial{k_{ef}}/\partial{q}\right)  \cdot \left(\partial{q}/\partial{X_i}\right)
\label{q}
\end{equation} 
The sensitivity coefficient with the restoration of $K_{ef}$ is 
\begin{equation}
 S(F_k|X_i)=\partial{F_k}/\partial{X_i} - \left(\partial{k_{ef}}/\partial{q}\right) \cdot  \left(\partial{q}/\partial{X_i}\right)
\label{S(K_{ef}}
\end{equation} 
In the discussed ephemerides problem, the variable parameters of the planetary system are subjected to   constraints supposed to be embedded into the model. They arise from conservation laws: the constancy of total energy $E$ (the symmetry of time translation that is, the balance of potential and kinetic energy), the constancy of angular momentum $L$ (the symmetry of 3-space isotropy that is, the absence of torque), and the constancy of linear momentum $P$ (the inertial system property that is, the absence of acceleration of a test particle at the barycenter). They should be considered similarly to the above example of neutron balance conservation. 

The existing practice of the ephemerides production suggests that every calculation of ephemerides should start with the``standard'' multi-digit initial values at a certain instant of time. From the statistical viewpoint, such a procedure makes a strong influence on the parameters' adjustment. The {\em a priori} data become subjected to the systematical off-set what results in diminishing of significance of new observational data.  This fact partly explains why results of calculations conducted in different ephemerides centers look stable and similar. It gives a false expression of a very high precision in descriptions of the Solar planetary system while the real precision remains unknown.

The presented method suggests that the initial data should be used once at a starting time. After that, the deviations from the once set  {\em a priori} database are calculated upon arrival of every new portion of observational data. A new precision is determined from the corresponding  {\em a posteriori} covariance matrices.

\subsection{Concluding remarks}

We described the method of observational data treatment based on a rigorous statistical theory. The reason for its presentation here is that in the existing practice of treatment of astronomical observation data the terms are used such as ``precision'', ``least square'',  ``fitting'', ``sensitivity'' and so forth, from the vacabluary of statistical theories. However, from what we lerned from literature related to the problem, those terms, as they are used in reality, have nothing to do with the statistical theory.

So far, we consider the Newtonian $N$-body model. An introduction of GR relativistic effects  inevitably leads to complications of the problem formulation for, at least, two reasons: a) such quantities as potential and kinetic energies are out of the GR arsenal; b) the exact GR $N$-body solution, which would have the Newtonian limit, does not exist. That is why we are critical of the PPN concept and oppose its introduction in the ephemerides systems. The basic relativistic effects are known without their GR connections and the corresponding corrections can be easily introduced.

There must be a few levels of parameter categorization with regard to their influence. The first level:  fundamental constants $G$, $c_0$ (in principle, they can be considered the exact numbers). The next level are basic physical characteristics such as point masses of the Sun and planets (the $N$ body problem). Further level would deal with most numerous and shaky parameters responsible for the tidal forces due to body size, composition and rotation, space debris, as well. Finally, one can think about relativistic corrections of the model but no new parameters are needed (except for the PPN case). It is important for researchers to have an opportunity to handle the parameters by their levels (to be able ``to turn'' them ``on'' and ``off'', for example, to compare results of Newtonian model with the one including relativistic corrections). Because parameters are characterized by a variety of measuring units, it is very convenient to formulate the problem in a dimensionless form.
 
The problem of initial conditions is emphasized in the context of ephemerides adjustment. In practical numerical calculations of ephemerides, several different files of ``standard'' initial data with a high number of significant digits are used, which are special parts of the systems. When new observational data became available for further testing and correcting the ephemerides data base, the same initial data are often used. It is explained why, from the statistical point of view, such practice is inconsistent with the statistical approach.  Renewable prior data files must launch the system from any instant $T$. 

Our main reason for discussions of the role of statistical approach to astronomical observation treatment is the fact that the GR effect of the Mercury's perihelion advance is claimed to be successfully confirmed. In the present work, the results of our analysis of the problem show that the claimed confirmation is not true. Astronomers have to bear some responsibility for their practice of observation treatment.

\end{document}